\documentclass[sigconf,authorversion,nonacm]{acmart}

\newcommand{\bfX}{\mathbf{X}}
\newcommand{\bfZ}{\mathbf{Z}}

\newcommand{\clG}{\mathcal{G}}

\usepackage{algorithm}
\usepackage{algpseudocode}
\algrenewcommand{\algorithmiccomment}[1]{\textcolor{gray}{$\sharp$ #1}}

\AtBeginDocument{%
  \providecommand\BibTeX{{%
    \normalfont B\kern-0.5em{\scshape i\kern-0.25em b}\kern-0.8em\TeX}}}

\setcopyright{acmcopyright}
\copyrightyear{2023}
\acmYear{2023}

\usepackage{xcolor}
\usepackage{comment}
\acmConference{When Causal Inference meets Statistical Analysis}{April 17--21,
  2023}{Paris, France}

\begin{document}

\title{Causal discovery for time series with constraint-based model and PMIME measure}

\author{Antonin Arsac}
\email{antonin.arsac@cea.fr}
\affiliation{%
  \institution{\textit{Université Paris-Saclay, CEA LIST,}}
  \city{Palaiseau}
  \country{France}
}
\author{Aurore Lomet}

\email{aurore.lomet@cea.fr}
\affiliation{%
  \institution{\textit{Université Paris-Saclay, CEA LIST,}}
  \city{Palaiseau}
  \country{France}
}

\author{Jean-Philippe Poli}

\email{jean-philippe.poli@cea.fr}
\affiliation{%
  \institution{\textit{Université Paris-Saclay, CEA LIST,}}
  \city{Palaiseau}
  \country{France}
}

\begin{abstract} 

Causality defines the relationship between cause and effect. In multivariate time series field, this notion allows to characterize the links between several time series considering temporal lags. These phenomena are particularly important in medicine to analyze the effect of a drug for example, in manufacturing to detect the causes of an anomaly in a complex system or in social sciences... Most of the time, studying these complex systems is made through correlation only. But correlation can lead to spurious relationships. To circumvent this problem, we present in this paper a novel approach for discovering causality in time series data that combines a causal discovery algorithm with an information theoretic-based measure. Hence the proposed method allows inferring both linear and non-linear relationships and building the underlying causal graph. We evaluate the performance of our approach on several simulated data sets, showing promising results.

\end{abstract}


\maketitle

\section{Introduction} \label{introduction}

Causality plays a key role in human understanding of the world. The last two decades have seen the development of formal causal models such as those proposed by Pearl and Halpern \cite{pearl2009causal,halpern2008causes}, that allow to take into account the distinction between causality and spurious correlation. Based on those developments, our work focuses 
on the expression of causality in temporal varying systems.

Time series can be found in many different fields such as medicine (e.g., neural sciences~\cite{sykacek2001bayesian} ), economy (e.g. econometrics, market studies~\cite{brodersen2015inferring}) or earth sciences (e.g., weather forecasting, climate sciences~\cite{runge2019inferring}). 
Several methods have been developed or improved for multivariate time series analysis. Those methods are generally 
based on the ability of algorithms to make links between data. Our aim is thus to implement a method able to find those links to build a causal graph from multivariate time series. In such graphs, each node corresponds to a variable at a given time. The presence of an edge between two nodes shows a direct dependency while no edge stands for independence or conditional independence \cite{spirtes2000causation}.

Once the relationships between variables of a system are found, reasoning on the basis of these links becomes possible, e.g. to  
predict the effect of an advertising campaign on the market. However, a great number of models are based on assumptions 
that may not fit reality, 
like linearity assumptions or a presupposed data distribution. To circumvent those problems, we aim at inferring causal relationships between multivariate time series, with as less assumptions as possible.

We develop a method that addresses the problem of causality for multivariate time series with few assumptions. It consists of merging a causal discovery algorithm, the PC algorithm \cite{spirtes2000causation}, with an information theoretic-based causal inference measure, the Partial Mutual Information from Mixed Embedding \cite{kugiumtzis2013direct} (PMIME). 
Based on information theory, the PMIME allows 
to limit assumptions on the data but also on the links between time series. With this measure, the PC algorithm gives causal relationships among multivariate time series, by representing the causality with a Directed Acyclic Graph (DAG).\\[2mm] 
This paper is structured as follows: section II introduces necessary notions to deal with causal discovery in time series and related works; section III focuses on the method we propose and we present results in section IV. Finally, we conclude and discuss our method and results in section V. 

\section*{Notations}
In the following, $X^0,...,X^g$ represent random variables, $X^i$ is the $i^{th}$ time series and $ X_t^i$ the $i^{th}$ time series at time $t$. Additionally, $X,\, Y,\, Z$ are random variables where $X$ is often the studied potential cause, $Y$ is its effect and $Z$ is a (potentially multivariate) conditional set. A multivariate process is written $\bfX = (\bfX_t, \bfX_{t-1}...)$. Let $n$ be the number of observations of a multivariate process of size $g$ at time $t$, $\bfX_t = \{x_t^0, x_t^1,..., x_t^g\} $ and $\bfX_t^- = (\bfX_{t-1},\bfX_{t-2},...)$ a multivariate process at all time before $t$. A lag is noted $\tau$ and the maximal lag considered is $\tau_{max}$. If $\clG$ is a causal graph, $X$-$Y$ denotes a link between $X$ and $Y$ and $X$ is called a neighbor of $Y$ in $\clG$ ($X \in adj(Y, \clG)$). If there is an arrow from $X$ to $Y$ ($X \rightarrow Y$), then $X$ is a parent of $Y$ in $\clG$ ($X\in Par(Y,\clG)$).

\section{Causality for time series} \label{causality_TS}

Causality between time series can be represented by causal Bayesian networks (CBN). 
 CBN is a class of graphical models that allows a probabilistic representation of random variables using graphs. The \textit{temporal priority} principle, which states that a cause precedes its effects, is useful in this frame. Indeed, this concept makes causality asymmetrical in time and will be of great help for orienting a causal relation in a graph, when a cause is already known.
With time-dependent variables, causality becomes temporal and a causal relation is a \textit{lagged causal relation} if a variable at time $t-\tau$ causes another variable at time $t$. If the cause appears at the same time as its effect, the relation is named an $instantaneous$ (or $contemporaneous$) cause.

Several types of graphs are used to represent causality for multivariate time series. 
\begin{figure}[h]
    \centering
    \includegraphics[width = 0.45\textwidth]{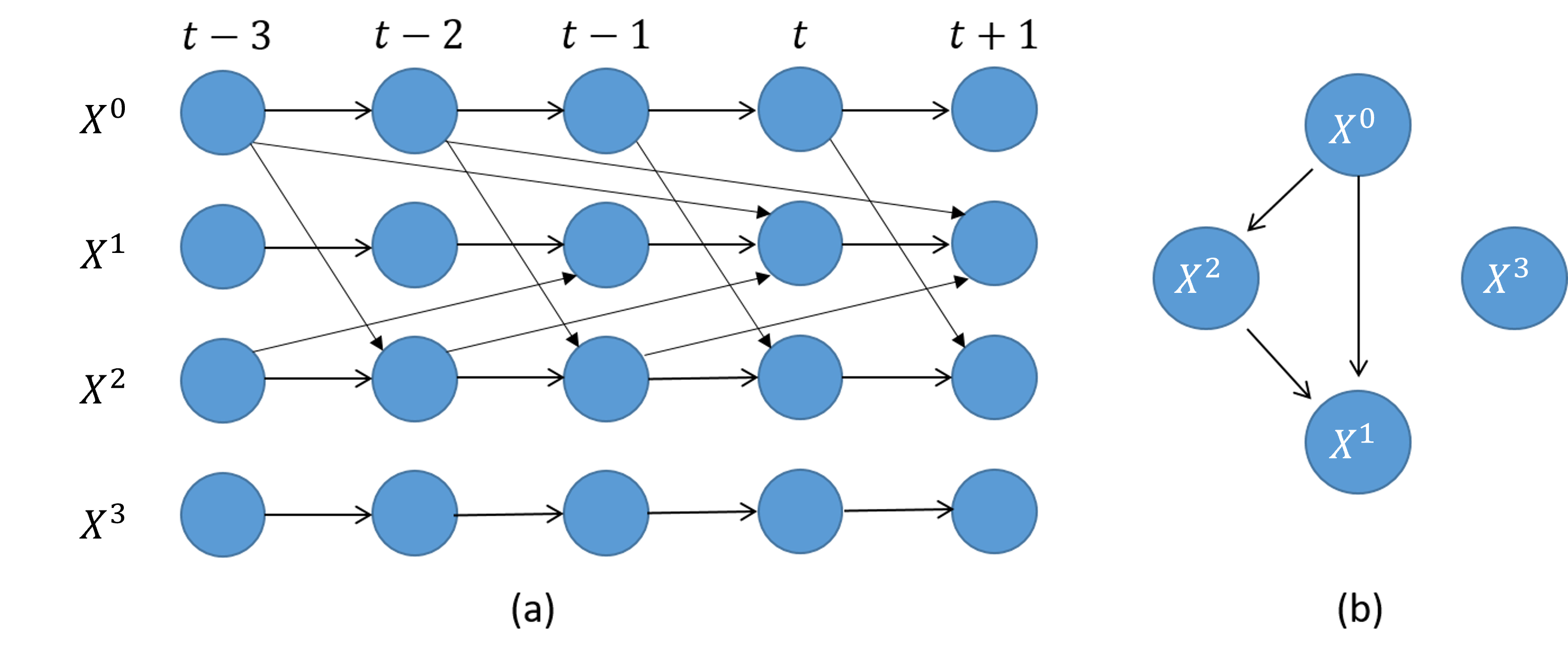}
    \caption{On the left (a), a window causal graph and its corresponding summary causal graph on the right (b).}
    \label{fig:graphs}
\end{figure}
Full-time causal graphs show all dependencies between every variables in the graph, at all time. Such a graph is difficult to represent in its entirety. Thus, under the assumption of causal stationarity \cite{runge2018causal}, stating that all causal relationships remain constant in direction throughout time, it can reduce 
to a window causal graph. 
This representation only provides a temporal section of the full time causal graph, after having selected a maximal time lag $\tau_{max}$ in the past and a maximal time ahead. The last well-known way to represent multivariate time series is the summary causal graph 
where all causal relationships in the graph are shown by keeping only variables and edges between them.
Thus, in this type of graph, the lag between a cause $X$ and its effect $Y$ is not illustrated. However, the summary causal graph has the advantage of being less impacted by noisy data, as it makes less calculations. Figure \ref{fig:graphs} presents a window causal graph on the left and its corresponding summary causal graph on the right.

Under specific assumptions (presented in  \cite{spirtes2000causation,peters2017elements}), those graphs encode conditional independence, leading to Directed Acyclic Graphs (DAGs). In a DAG $\clG$, arrows connecting two nodes stand for direct dependency, while no arrow between two nodes shows either independence or conditional independence. However, an issue that may occur is that two different DAGs can encode the same dependencies. In this context, we say that those graphs belong to the same Markov equivalence class \cite{verma1991equivalence}.\\

Searching causal relationships in data can be seen as a statistical estimation of parameters describing a causal structure. Such a problem is categorized under the terms of \textit{causal discovery}, which aims at using observational data to analyse and identify properties or causal relationships of a system. When dealing with temporally varying systems, causality is generally associated with Granger causality \cite{granger1969investigating}. It states that variable $A$ Granger-causes variable $B$ if the prediction of the future of $B$ is improved when the knowledge of the past of $A$ becomes available.

 A major restriction of this model is that it focuses on linear relations. Another issue with Granger causality is that by definition, it does not account for instantaneous effects. Despite this, there have been various 
 
 formulations of Granger causality and methods have been 
 developed to express this notion 
 on the basis of observed time series. In a typical approach to check for Granger causality, two models are fitted: one full, composed of the past from every time series involved and the other one, reduced, that does not include the past of the investigated time series. The two models are then compared with respect to a measure of prediction. 
 If the full model performs better than the reduced one, then the causality is inferred. This approach is called Parwise Granger Causality (PWGC) and extensions exist to process multivariate time series \cite{chen2004analyzing, barrett2010multivariate}. Other methods have been developed and are grouped under three families: \textit{functional causal models} (FCMs), \textit{score-based approaches} and \textit{constraint-based approaches}.
 
 The FCMs are based on Structural Equation Models \cite{pearl2009causal}. 
 Their objective is to make a correspondence between a graph $\mathcal G$ and a system of equations in which each variable is expressed in terms of its direct causes and an additional noise. In the frame of time series data, 
 some of the most popular algorithms are the VarLiNGAM \cite{hyvarinen2010estimation} and TiMINo \cite{peters2013causal}. The first one is an extension of LiNGAM \cite{shimizu2006linear} for time series, with the use of auto-regressive vectors, while the second one discovers causality through statistical tests to look for independence between residuals and noise of a certain fitted time series model. Both methods are originally developed considering simplifying assumptions on the relationships between variables.
 
 Score-based models start with an empty graph and add (or remove) needed (or unnecessary respectively) edges in an iterative pattern. They consist in searching for graphs maximizing the goodness of fit to the data distribution. For time series, a recent algorithm, DYNOTEARS \cite{pamfil2020dynotears} considers dynamic Bayesian networks in which variables are time series. The major issue is that this approach does not guarantee that the inferred graph belongs to an equivalence class. 
 
 Lastly, constraint-based approaches try to efficiently search for graphs belonging to a Markov equivalent class that fit the most the set of conditional independence relations found in the data. The two main algorithms that hold in constraint-based methods are namely the 
 PC algorithm 
 and the Fast Causal Inference algorithm~\cite{spirtes2000causation}. The first one makes the assumption of causal sufficiency while the latter does not. We will further explore the PC algorithm. 
 
 PC algorithm starts with a complete undirected graph. It ends with an oriented graph where only edges that connect two nodes that have a causal relation are kept. This algorithm is composed of three different parts. The first one intends to estimate the skeleton of the graph by testing (conditional) independence between variables. If two variables are independent or conditionally independent, then it removes the edge between them. The second one identifies the $v$-structures in the graph and directs the concerned edges, from the separation sets found in the first phase. The last one makes use of the knowledge brought by the previous steps to finish directing the graph. PC algorithm does not ever return a DAG, but a complete partially DAG where some edges may remain unoriented, which is sure to belong to the right Markov equivalence class.

For time series, several methods have been proposed based on the PC algorithm, such as oCSE \cite{sun2015causal} or PCMCI \cite{runge2019detecting} and its extensions, using the (Conditional) Mutual Information ((C)MI) to measure dependencies between variables. Indeed, the information theory framework allows to limit assumptions on the data distribution as well as on the relationships between them. As such, some information theoretic based measures have been developed specifically to process temporal variables such as the Transfer Entropy \cite{schreiber2000measuring} (TE), generalized as the partial Transfer Entropy \cite{vakorin2009confounding} (PTE), or the Partial Mutual Information from Mixed Embedding \cite{kugiumtzis2013direct} (PMIME).

PMIME is an asymmetric and non-parametric measure designed to detect direct couplings in time series. It is derived from an embedding scheme based on a selection criterion, the conditional mutual information. In the multivariate case, to assess whether a variable $X$ leads a variable $Y$, conditional on a set of variables $Z=\{Z^0,Z^1,\dots,Z^{g-2}\}$, it builds iteratively an embedding vector $\textbf{w}$ from the lagged components extracted from $(X,Y,Z)$ that explains the best the future of $Y$, noted $Y_t^T = (Y_{t+1},...,Y_{t+T})$. Each iteration is called an embedding cycle and uses a stopping criterion to accept or reject a component. A component is accepted if the information it brings strictly increases the information already contained in the embedding vector.

Hence $\textbf w$ is formed from $k$ lagged variables, selected by the CMI and can be decomposed as $\textbf w_t = (w_t^x,w_t^y,w_t^Z)$, where $w_t^x$ are the components of $X$ selected in the process, $w_t^y$ those from $Y$ and all the remaining ones are denoted as $w_t^Z$, see more details on the embedding process in \cite{kugiumtzis2013direct, vlachos2022phase}.\\ We then quantify the causal effect from $X$ to $Y$, conditional on $Z$ by : $$R_{X\rightarrow Y |Z} = \dfrac{I(Y_t^T,\textbf w_t^x|\textbf w_t^y,\textbf w_t^Z)}{I(Y_t^T;\textbf w_t)}.$$

The embedding process alone could be a measure of causality. Indeed, if $w_t^x$ is empty, it means that $X$ has no influence on $Y$, which translates in the measure $R$: if $w_t^X$ is empty, then $R$ equals $0$. Furthermore, this measure is bounded between $0$ and $1$, $0$ means independence and $1$ means that the future of $Y$ is totally driven by $X$.

\section{Proposed approach} \label{pcpmime}
\subsection{Motivations and assumptions}

In this work, we consider multivariate 
To not restrict real world assumptions, we focus on variables holding linear and non-linear, possibly lagged relationships. Additionally, time series data are suppose not to respect any particular probabilistic distribution model. Lastly, we make the assumption that every causes of each effect are observed, known as causal sufficiency. 
As such, most of the state-of-the-art algorithms are not adapted to our framework. In a recent survey \cite{assaad2022survey}, 
all algorithms, except PCMCI, tsFCI \cite{entner2010causal} and oCSE, work under linear relationships or with a particular data 
distribution 
(e.g. Gaussian or vector auto-regressive models). In the remaining ones, oCSE considers only a maximum lag of 1, which might not be sufficient to fully grasp temporal relationships in real-world time series. 
Based on the FCI algorithm, tsFCI aims at discovering hidden confounders, which is not in line with our causal sufficiency assumption. Therefore, the methods that meet our requirements the more are PCMCI and its derivatives. Those are constraint-based methods so they require a suited causality inference measure. One issue is that PCMCI uses either Partial Correlation or Mutual Information to find independence and conditional independence. Partial Correlation is a measure that makes the assumption of linear links and is therefore not suitable to our framework and (Conditional) Mutual Information alone might not be able to fully detect lagged relationships. Another issue is that PCMCI uses a window causal graph that can be more sensible to noise and costly due to the need to identify all causal relations at all lags in the window.

To address these problems, a combination of the PC algorithm, a constraint-based method and the PMIME measure is further developed to infer causality in time series data in our framework. Indeed, the PMIME measure is adapted to quantify 
temporal links with 
lagged relations. 
 It is built for no particular probability distribution of the system, can process linear and non-linear links and has few free parameters to adjust.\\
The Partial Transfer Entropy also respects the precedent properties but the PMIME has been shown to be more efficient than the PTE in the frame of nonlinear systems \cite{papana2013simulation}. Moreover, the fact that the PMIME measure is bounded has two advantages: firstly, it does not require additional significance test; secondly, this bounded score makes it easier to interpret the results.

The PC-PMIME method keeps some important assumptions. The first major one is the causal sufficiency due to PC algorithm limits: it cannot discover hidden confounders or selection bias variables. Additionally, PMIME requires stationary time series. Lastly, the estimation of the Entropy and through this, of the CMI, requires large time series. Also, due to the form of the PMIME measure, the PC-PMIME algorithm does not compute auto-correlation. Indeed, in PMIME, if $X = Y$, then the variables of $X$ are the same as those from $Y$ in the embedding vector, thus $w^X = w^Y$ and $R_{X\rightarrow Y} = R_{Y_t^- \rightarrow Y} = 0$. The measure of auto-correlation could be integrated in future work.

\subsection{PC-PMIME algorithm}
\begin{algorithm}
\caption{\textit{PC-PMIME}}\label{pcpmime_alg}
\begin{algorithmic}[1]
\Require $n$ observations of $X^0, \dots, X^g$, other parameters of PMIME 
\Ensure $\clG$, the estimated graph
\State Create a complete graph $\clG$ with $g$ nodes
\State Initialize $rm\_e$  the list of removed edge
\For{each permutation $(X^i,X^j)$ of nodes from  $V$} 
\State Compute the PMIME measure $R$ with $X^j$ and $X^i$ without conditioning
\State Affect $R$ to the edge $X^j \rightarrow X^i$
\If{$R \approx 0$}
\State add the edge between $X^i$ and $X^j$ into the list of removed edges
\EndIf
\EndFor
\State Remove all edges in $\clG$ contained in $rm\_e$
\State Initialize $l = 1$, $process$ = True
\While{ $process$ is True}
\State Set  $process$ to False
\State Reinitialize $rm\_e$ = [ ] the list of removed edges
\For{each $(X^i,X^j)$ permutations of nodes from  $V$} 
\State $adj\_set$ =  $Par(X^i,\clG)\backslash X^j$ the list of predecessors of $X^i$, without $X^j$
\If{$Card(adj\_set) \geq l$}
\For{each combination $\bfZ$ of nodes from $adj\_set$ of size $l$} 
\State Compute R, the PMIME measure of $X^j \rightarrow X^i |\bfZ$
\State Affect R to the edge between $X^j$ and $X^i$
\If{R is near 0}
\State Add the edge $X^j \rightarrow X^i$ into the list of removed edges
\State \textbf{Break}
\EndIf
\EndFor
\State Set  $process$ to True
\EndIf
\EndFor
\State Remove all edges in $\clG$ contained in $rm\_e$
\EndWhile
\State Orient all edges $X^j \rightarrow X^i,  \, \forall(X^i,X^j)\in \clG$ if $R_{X^j\rightarrow X^i} > 0 $\\
\Return The causal graph $\clG$
\end{algorithmic}
\end{algorithm}
To find the causal structure in time series data, a causal discovery algorithm, PC algorithm, is merged with the non-parametric measure PMIME. Only the first phase of the PC algorithm is used: starting from a full connected graph, it finds the skeleton by successively testing every edges between each node. For instance, an edge between $X$ and $Y$ is removed if $R_{X \rightarrow Y}=0$, where $R$ is the PMIME measure. When all edges have been tested and some have been removed, for the remaining edges, it continues by checking if the two connected nodes are conditionally independent. The conditioning set (or separation set) is first composed of one additional variable, connected to $X$ or $Y$ and its size increases until conditional independence is found or until all edges linked to $X$ and $Y$ have been tested. As the PMIME is asymmetric, the algorithm tests for both directions, from $X$ to $Y$ and from $Y$ to $X$, to make sure it does not make spurious links. In our implementation, an edge is not removed as soon as $R = 0$, but instead when the algorithm has tested all edges for one size of the conditioning set. This is known as the PC-stable method \cite{colombo2014order} that allows to avoid PC to be order biased.

The algorithm is described in Algorithm \ref{pcpmime_alg}. PC-PMIME\footnote{PMIME and PC-PMIME are implemented in \url{https://github.com/AArsac/CD_for_TS_with_CBM_and_PMIME} } takes as inputs the data ($g$ time series of length $n$), a maximal lag $\tau_{max}$, $k$ the number of nearest neighbors for the estimation of the CMI and $A$, the value of the stopping criterion in the building of the embedding vector. The algorithm returns $\clG$, the estimated oriented causal graph. We observe that in general, $R$ is not equal to $0$ in case of independence, but is more around $10^{-15}$, due to estimation errors. So we consider that there is independence when $R$ is close to $0$ (we note $R \approx 0 \Leftrightarrow R < 10^{-10}$).\\
Although the PC-PMIME algorithm is not fitted to search for latent variables, the implementation is made in a way that if $R_{X^j \rightarrow X^i} > 0$ and $R_{X^i \rightarrow X^j} > 0$, then it leads to a double headed arrow ($X^j \leftrightarrow X^i$). That provides the information that those two variables are mutually correlated and that there is potentially a common confounder between those two.

\section{Experiments} \label{expe}

\begin{figure*}[h!]
\centering
\includegraphics[width=0.7\textwidth]{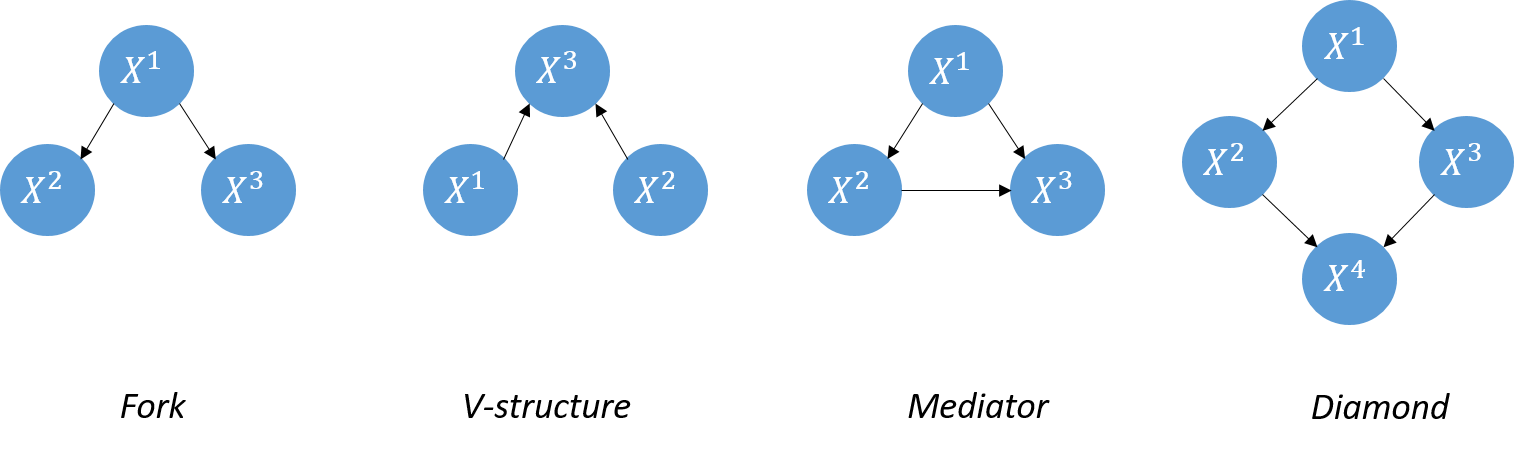}
\caption{Simulated basic causal structures}
\label{structures_causales}
\end{figure*}

Our implementation of PC-PMIME is tested on simulated data, from a recent survey \cite{assaad2022survey}. In this survey, the authors simulated basic causal structures often encountered, from time series. It contains a total of 5 different structures, each simulated 10 times over 4000 observations. From the five structures, only four are retained, as the last one contains latent variables. Those four are the \textit{Fork}, the $v-structure$, the $Mediator$, and the $Diamond$ structure, as shown in Figure \ref{structures_causales}. The Fork structure corresponds to a common confounder, the $v-$structure is implicit. Then, Mediator corresponds to a collider with one of its parent causing the other. Finally, the diamond structure is a common confounder leading to a $v-$structure. The data are simulated with linear relations for auto-correlation and nonlinear links between different time series, through simple nonlinear functions.\\

\begin{figure*}[h]
\centering
\begin{tabular}{ll}
\includegraphics[width=0.45\textwidth]{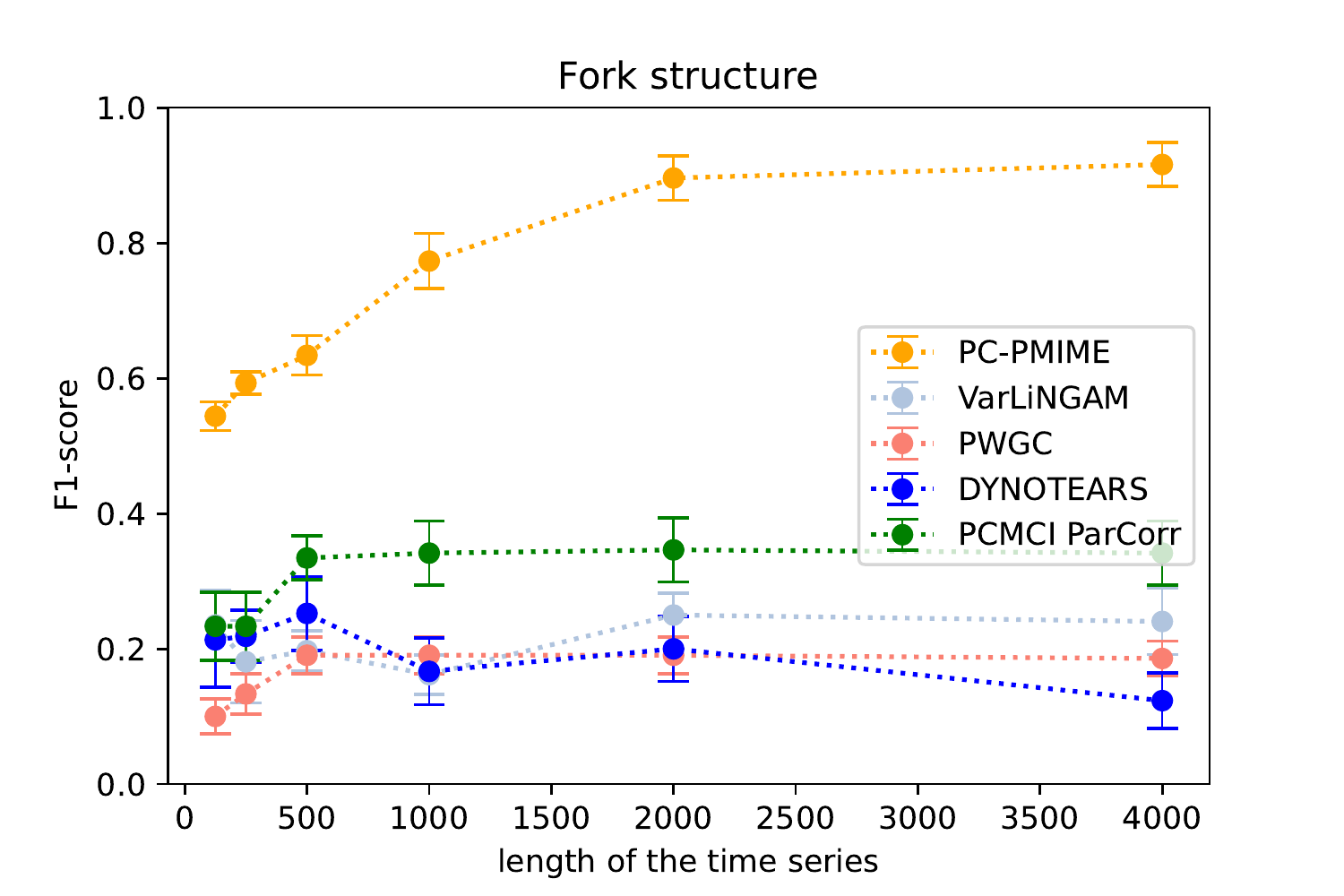} &
\includegraphics[width=0.45\textwidth]{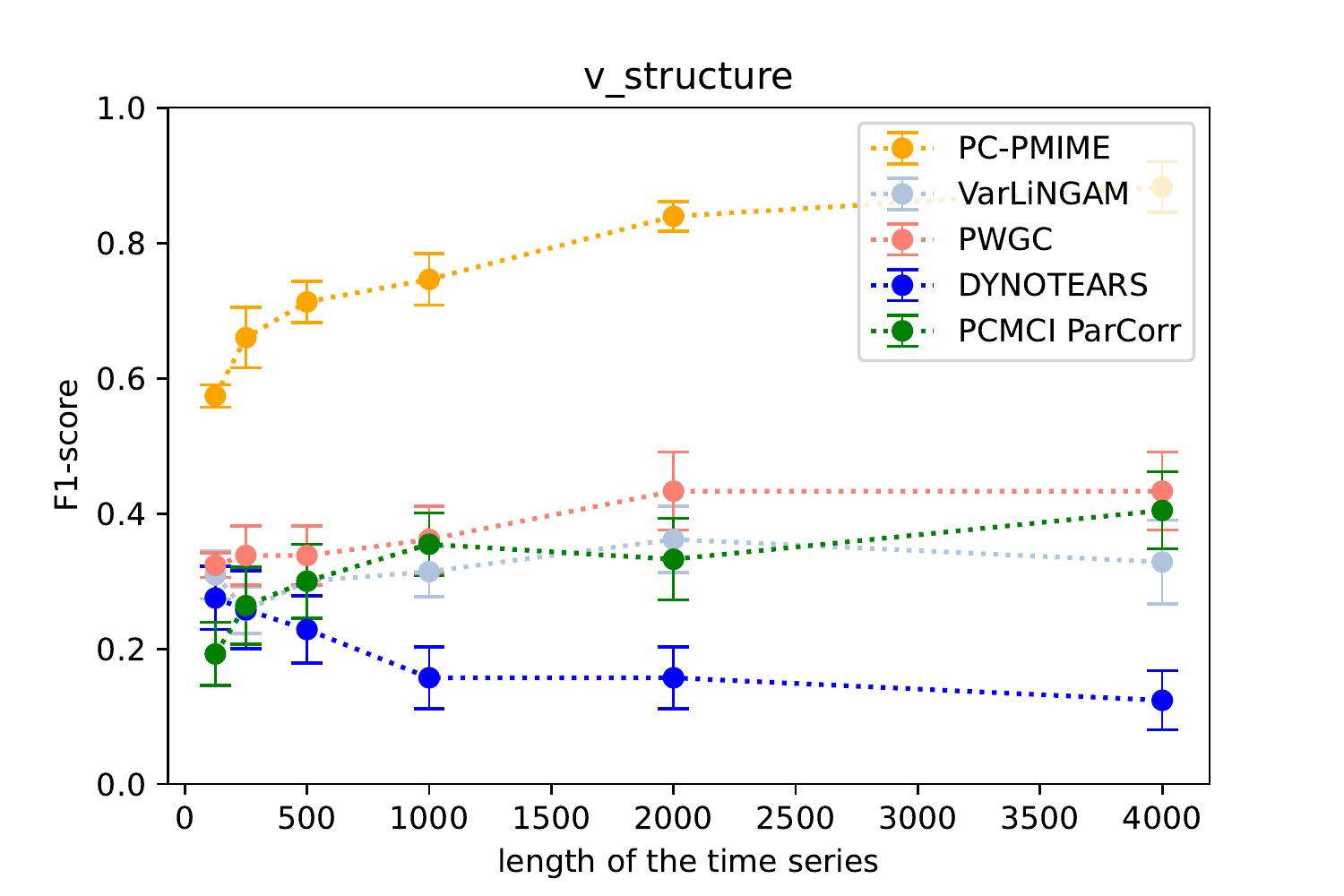} \\
\includegraphics[width=0.45\textwidth]{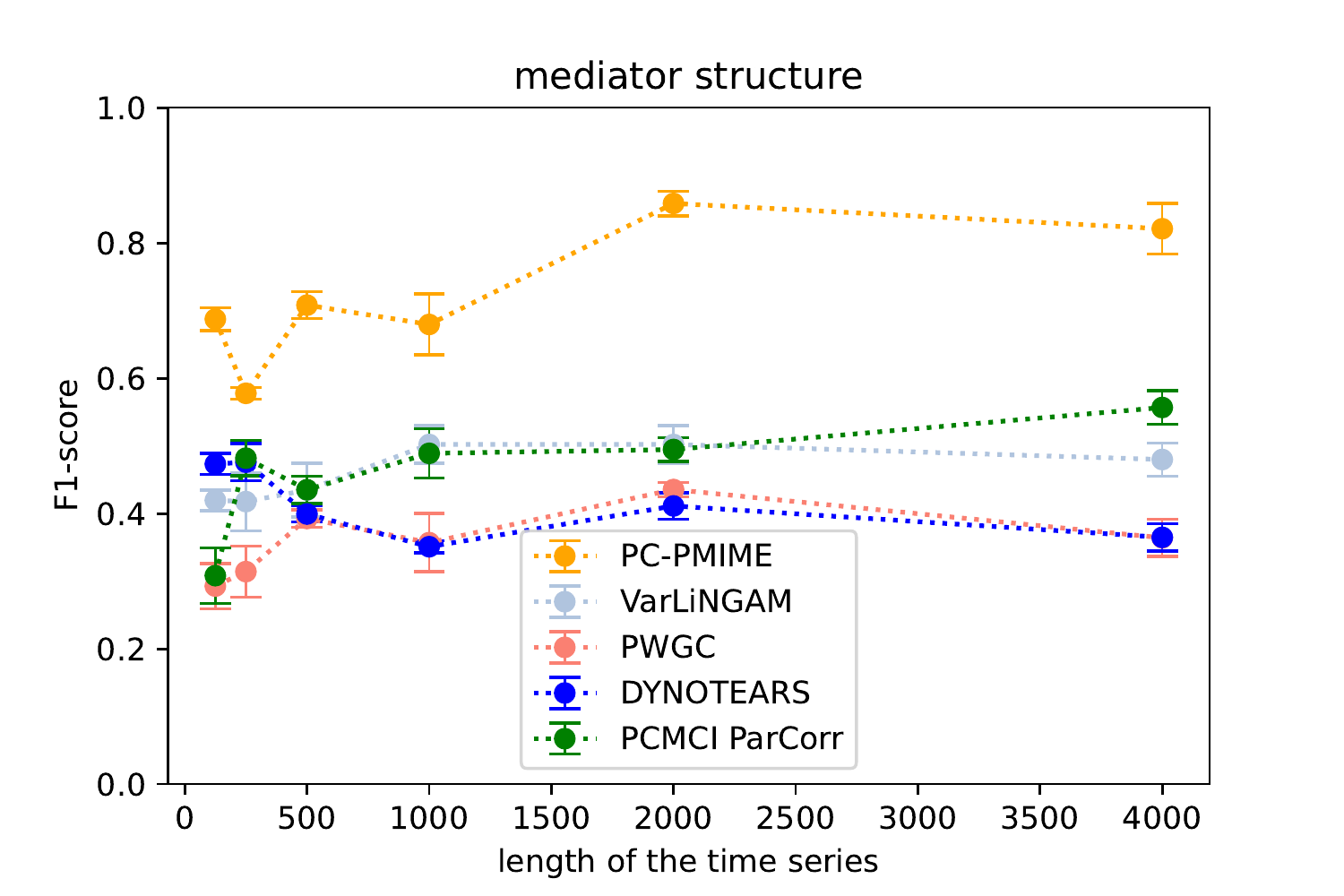} &
\includegraphics[width=0.45\textwidth]{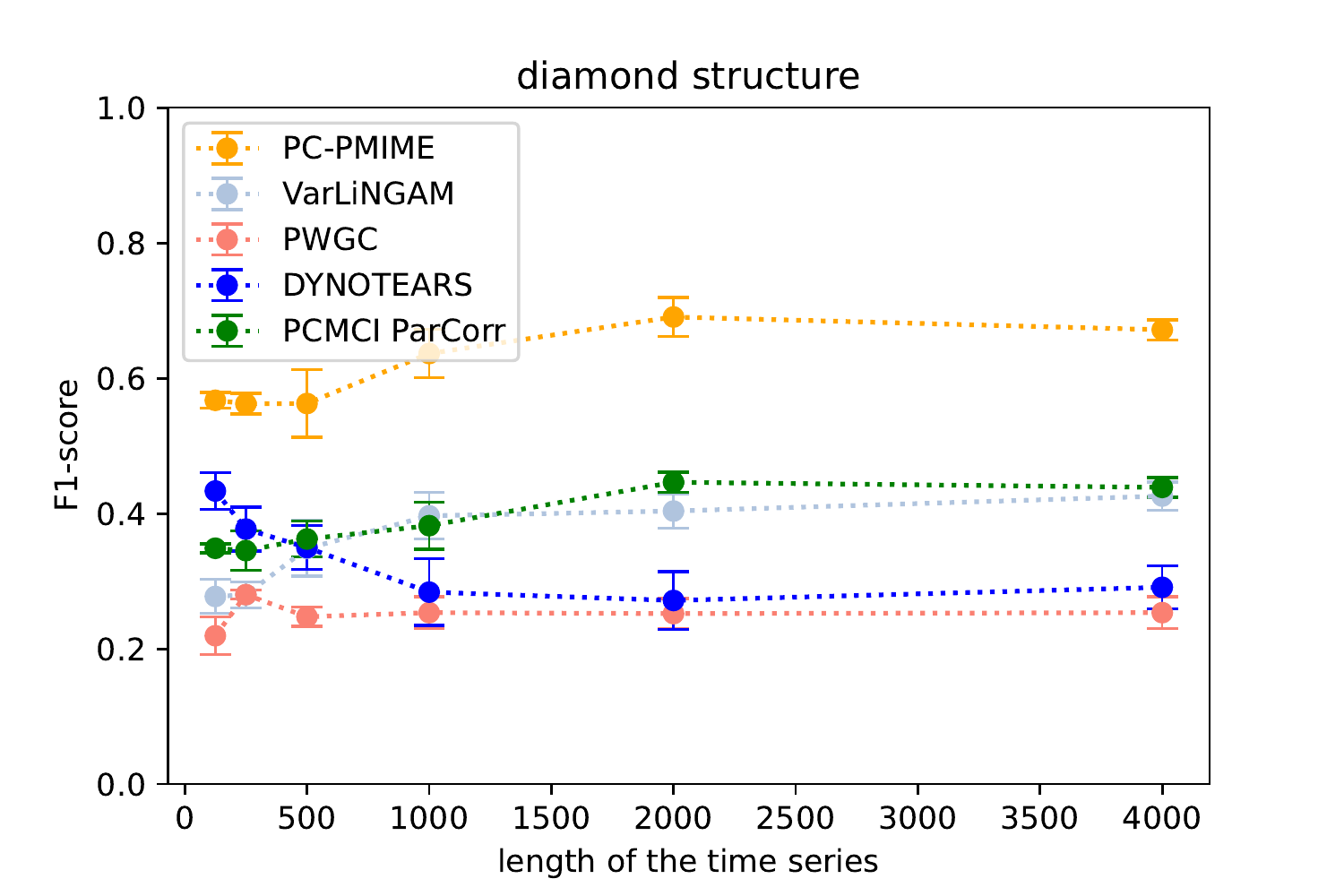} \\
\end{tabular}
\caption{$F1-$score of the 5 methods in function of $n$, the length of time series. Each graph corresponds to one basic causal structure. }
\label{results1}
\end{figure*}
On each structure, PC-PMIME as well as four other methods are run. Those methods are the Pairwise Granger Causality (PWGC\footnote{PWGC algorithm can be found on \url{https://www.statsmodels.org/stable/generated/statsmodels.tsa.stattools.grangercausalitytests.html}.}), VarLiNGAM\footnote{VarLiNGAM is implemented in \url{https://github.com/cdt15/lingam}}, DYNOTEARS\footnote{Dynotears can be found on https://github.com/quantumblacklabs/causalnex} and PCMCI\footnote{PCMCI is part of the Tigramite library https://github.com/jakobrunge/tigramite} with Partial correlation as a measure of conditional independence. Those are only baselines to compare to, other algorithms exist but are not tested here. For each algorithm, including PC-PMIME, the maximal lag is defined at $\tau_{max} = 3 $, selected empirically on these datasets. For PWGC, the statistical test to compare the full model and the restricted is the $F-$test, with a significance level set at $\alpha = 0.03$. 
VarLiNGAM uses a Lasso penalization for the estimation of the structural VAR model, whose parameters are selected by the Bayesian Information Criterion (BIC). The parametrization of DYNOTEARS is done with the recommended values $\lambda_w = 0.05 = \lambda_a$, and $w\_threshold = 0.01$ by \cite{pamfil2020dynotears}. Lastly, the significance value for the Partial Correlation measure in PCMCI is set to $\alpha = 0.03$. About PC-PMIME, for the estimation of the CMI, the number of nearest neighbors $k$ is set to $k = 0.01n$ (see \cite{runge2018conditional, frenzel2007partial}). After several tests, it appears that if the stopping criterion of the embedding cycles in PMIME, $A$ is close to $0$, such as $A = 0.01$, it is too conservative, while if it is greater than $0.05$, it is too permissive. Thus, the stopping criterion is fixed to $A = 0.03$. 

To evaluate each method, we resort to the $F1-$score. The closer its value is to $1$, the more the method performs. In this framework, a ``True Positive'' value occurs if a link in the estimated graph is the same as a link in the true graph. Moreover, auto-correlation is not measured, as PC-PMIME does not integrate it. \\
For each structure, we run the algorithm on the $10$ different data sets, with different length of time series $n = \{125,250,500,1000,2000,4000\}$. We then compute the mean and the standard deviation of the $F1-$score for each value of $n$. The scores shown in figure \ref{results1} are different from those in the survey \cite{assaad2022survey}. Indeed, accounting for auto-correlation in the score increases its value. For example, if we consider few nodes, thus few edges, accounting for auto-correlation as a true value augments significantly the score. Therefore, auto-correlation is still computed in the different methods but is not taken into account in the score. The focus is done on the graph discovery.

The scores obtained by the different methods on the simulated data are shown in figure \ref{results1}. PC-PMIME seems to perform really well as opposed to the others in each structure. Indeed, the $F1-$score reaches a mean of 0.9 for high values of $n$ on each structure, except on the diamond structure, while the four other algorithms barely reach 0.5. We also note that the standard deviation is really small. However, our method is not stable for smaller sizes of time series (under $n = 1000$). This is due to the PMIME measure that is based on an estimation of the mutual information. Actually, the mutual information and more precisely, the $knn$ estimation of the entropy, is robust asymptotically, hence behaves well for larger size of data. \\
\vspace{-1cm}
\section{Conclusion and future work} \label{discussions}

In this paper, we present a method to infer causal relationships between multivariate time series, making few assumptions about the data. As such, PC-PMIME is presented as a method to build a causal graph from nonlinear multivariate time series. It provides promising results on simulated data with no latent variables. However, it still contains several limitations that can be removed such as making a better orientation phase of edges, computing instantaneous relationships or taking into account hidden common confounders and testing on real data. 

In the current PC-PMIME algorithm, edges are only oriented from the result of the asymmetrical PMIME measure. However, it may not catch true causality and may also lead to spurious conclusions in the building of the causal graph. Thus, adapting the orientation rules of the PC algorithm to time series might be a better approach and should lead to an improved version of our method.

Then, to measure instantaneous relationships, the idea could be to work on a different type of causal graphs. In this study, summary causal graphs are used, but extended summary causal graphs might be more appropriate \cite{assaad2022discovery} to consider those relationships. With such graphs, the PMIME measure could infer lagged causal relationships and a simple measure of (conditional) dependence for instantaneous causes.
\begin{figure}[h!]
    \centering
    \includegraphics[width = 0.25\textwidth]{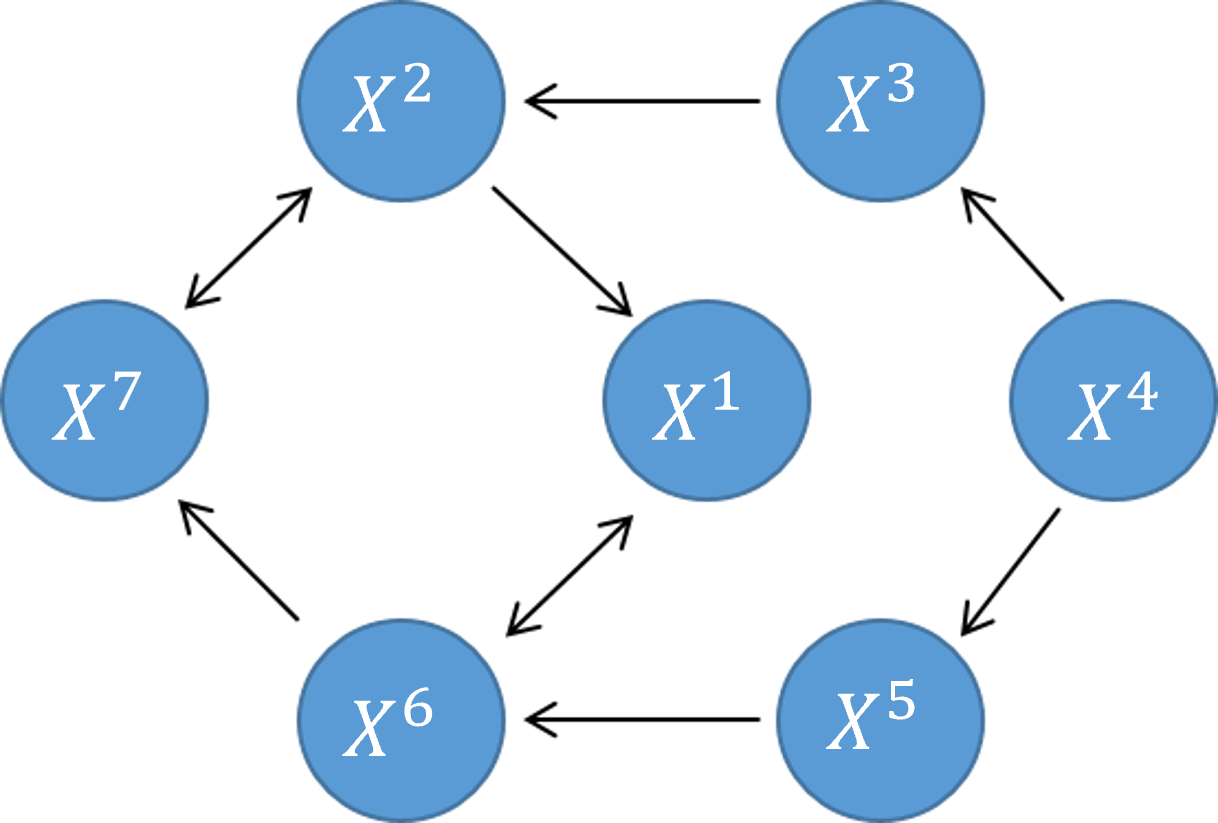}\\[1mm]
    \includegraphics[width = 0.44\textwidth]{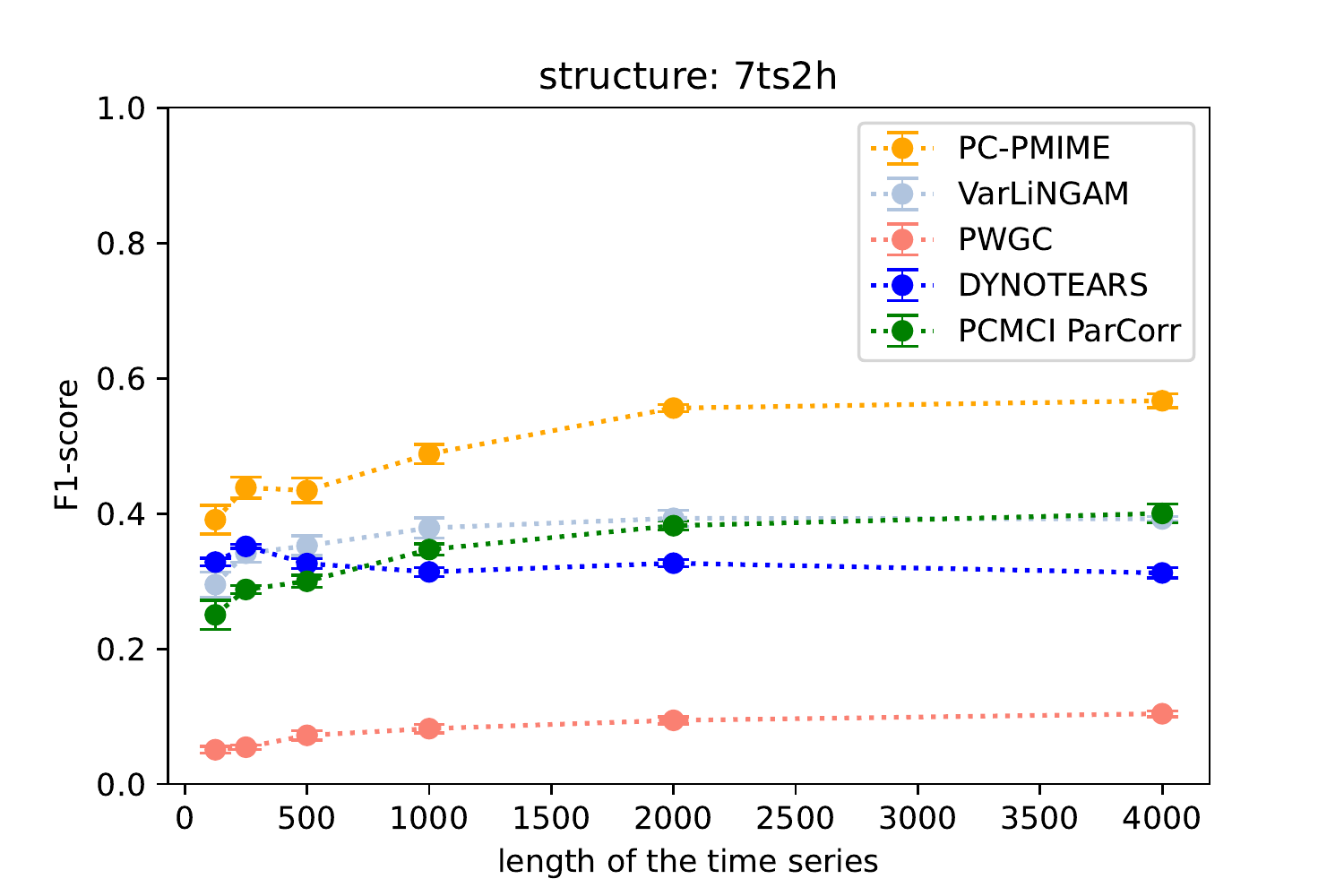}
    \caption{The first graph, above, is a summary causal graph of the structure simulated with $7$ time series, including $2$ latent variables, marked with double-headed arrows. Below, the $F1-$score of the 5 methods in function of $n$, on this structure.}
    \label{fig:7ts2h}
\end{figure}
Lastly, taking into account latent variables may be the highest challenge. The PC algorithm cannot be used anymore as it works under the assumption of causal sufficiency. To confirm this, we run the different methods on the last dataset proposed in \cite{assaad2022survey} that contains two hidden confounders, hence does not respect anymore the causal sufficiency constraint. The parameters for each method are the same as before. Figure \ref{fig:7ts2h} presents the true graph and the results obtained on these simulations. As expected, the scores obtained by the different algorithms are below those found on the other datasets with no latent variables. The one with the higher score is still the PC-PMIME algorithm, barely reaching a score of $0.6$. This is surely due to its ability to represent mutual correlations, as exposed at the end of section \ref{pcpmime}, but it is still not really satisfying and needs to be improved.

Thus, computing lagged variables requires the use of another algorithm, such as the FCI algorithm and adapting it to time series. The idea is then to merge an FCI-like algorithm with the PMIME measure in further works.

\bibliographystyle{unsrt}
\bibliography{References}

\end{document}